\documentclass[arial,11pt,a4paper]{article}
\usepackage{setspace}
\usepackage{textcomp}
\usepackage{mathcomp}
\usepackage{mhchem}
\usepackage{amssymb}
\usepackage{amsmath}
\usepackage{longtable}
\usepackage{upgreek}
\usepackage{fixltx2e}
\usepackage{wrapfig}
\usepackage{array}
\usepackage{subfig}
\usepackage{graphicx}
\usepackage{multirow}
\usepackage{sidecap}
\usepackage[top=2.5cm, bottom=2.5cm, left=2.5cm, right=2.5cm]{geometry}
\usepackage{hyperref}
\usepackage{parskip}
\usepackage{url}
\usepackage{natbib}
\usepackage{xfrac}
\usepackage{listings}
\usepackage{float}
\usepackage{amsthm}

\newtheorem{defi}{Definition}
\newtheorem{rem}{Remark}
\bibpunct{[}{]}{,}{s}{,}{,}

\DeclareGraphicsExtensions{.pdf,.png,.jpg,.tif}
\graphicspath{{graphics/}}

\begin{document}
\title{Modelling the burden caused by gene expression: an \emph{in silico} investigation into the interactions between synthetic gene circuits and their chassis cell}
\author{R. J. R. Algar, T. Ellis, G.-B. Stan}
\date{\today}




\maketitle




In this paper we motivate and develop a model of gene expression for the purpose of studying the interaction between synthetic gene circuits and the chassis cell within which they are inserted. This model focuses on the translational aspect of gene expression as this is where the literature suggests the crucial interaction between gene expression and shared resources lies.

\section{Background}

Numerous models have been proposed to capture the interactions between synthetic circuit and the host cell through shared resource pools \cite{scottinterdependence2010,perettimechanistically1986,perettisimulations1987,levycoordination2009,marchisiocomputational2008,klumppgrowth2009}.

Klumpp et al. \cite{klumppgrowth2009} explore the relationship between growth rate and a range of cellular metrics that influence gene expression such as cellular RNA levels and protein production rates. They alter growth rates using a range of 5 different growth media and show that there is a proportional relationship between growth rate and levels of RNA in the cell. In addition they show that there is no growth rate dependence for the translation rate in cells (defined as total cellular protein divided by total cellular RNA). They argue that cells use feedback mechanisms involving growth rate to auto regulate certain key processes and also suggest the possibility for designing circuits which utilise growth feedback in their behaviour (as seen in Tan et al. \cite{tanemergent2009}).

Mitarai et al. model the translational process and treat it as a traffic problem on which they perform stochastic simulations \cite{mitarairibosome2008}. They observe `traffic jams' of ribosomes on the mRNA where there are codons which are translated at a slower rate. They also report that codon usage impact the total number of ribosomes on a transcript. The effect of traffic jams and slow codons can be minimised using an `on ramp' of slower codons at the start of the transcript, as well as using a weaker RBS (though there is a threshold of RBS strength below which no additional benefit is gained). A ribosomal `cost' of protein production per transcript is also discussed.

There have been attempts to develop basic frameworks for predicting the behaviour of genetic parts when used in combination. Marchisio et al. \cite{marchisiocomputational2008} demonstrate a novel approach to this by modelling systems based on fluxes of cellular machinery such as polymerases, ribosomes, transcription factors and environmental signals \cite{marchisiocomputational2008}. This modelling framework was implemented within the ProMoT systems modelling and design tool and allows for individual parts to be composed into larger networks, something that is key to making synthetic biology design predictable and modular. The models used in this study are very simple and lack some key considerations, such as codon usage. There is also no provision of a methodology for characterising the individual parts to obtain the information required by the model to make predictions. However, an expansion of this approach may lead to an improved framework whereby predictions can be made about cell-circuit interactions.










\section{Full elongation model}

Experimental results reported in the literature seem to indicate that the major bottleneck in resource availability in cells expressing heterologous protein is greatest at the ribosomal level \cite{vindsynthesis1993,de_voshow2011}. Therefore we chose to build a model that focuses on ribosomal availability. In order to include the ability for coding regions to have different codon profiles and for ribosomal traffic jams \cite{chouclustered2004,basutraffic2007,mitarairibosome2008} to be able to be modelled it was necessary to go beyond `one step' translation models such as:

\begin{center}

\ce{Pol + DNA <=>[\alpha^1_+][\alpha^1_-] S}

\ce{S ->[\beta^1] Pol + DNA + RNA}

\ce{Rib + RNA <=>[\alpha^2_+][\alpha^2_-] L}

\ce{L ->[\beta^2] Rib + RNA + P}

\end{center}

where `Pol' is an RNA polymerase, `DNA' is a strand of DNA, `S' is a transcriptional complex of DNA and RNA polymerase, `RNA' is an mRNA transcript, `Rib' is a ribosomes, `L' is a translational complex of RNA and ribosomes and `P' is a protein.

To create a model that could capture key features, such as the ability for coding regions to have different codon profiles and the impact of codon profiles on the translation dynamics including ribosomal traffic jams, we had to look closely at the elongation process that occurs during mRNA translation. This is a complex process that consists of multiple steps every time the polypeptide chain is elongated and the ribosome moves along the mRNA transcript. Since obtaining the values for the parameters associated with these individual processes was not possible we collapsed them down in a way that each time the ribosome moved one codon along the transcript it was a single process. It is typically not possible to obtain these rates either. However we were able to roughly estimate these rates from typical values reported in the literature and roughly model how the differences in codon usage might be qualitatively reflected in the elongation rates used in our proposed model.

We derived our model using a random-walk approach, although a very similar model was derived elsewhere using more mechanistic and deterministic approaches \cite{macdonaldconcerning1969}. We confirmed that the model made sense by instantiating it with parameters found in the literature and making sure that the model outputs reflected what would be expected \emph{in vivo}. Subsequently we simulated what the effects of changing various control points such as promoter strength, transcript copy numbers, RBS strength or codon usage are for key metrics. Growth rate is not included in this model as the complexity of the interaction between resource availability and growth rate is still an open and debated question.

\subsection{Model derivation}

A model of translation was constructed where the movement of individual ribosomes is treated as a random walk which occurs based on the following assumptions:

\subsubsection{Assumptions}
\newtheorem{assumption_model}{Assumption}
\begin{assumption_model}
There is a fixed number of ribosomes R.
\end{assumption_model}
\begin{assumption_model}
There is a single species of transcripts of which there is a constant number M.
\end{assumption_model}
\begin{assumption_model}
Each transcript is identical and is of length L codons.
\end{assumption_model}
\begin{assumption_model}
Ribosomes can reversibly bind to the RBS of a transcript.
\end{assumption_model}
\begin{assumption_model}
Once elongation is initiated and a ribosome has moved to the first codon of the transcript, it must continue unidirectionally along the transcript until it reaches the stop codon.
\end{assumption_model}
\begin{assumption_model}
When a ribosome reaches the stop codon it will release it and become a `free ribosome' again and a protein will be produced.
\end{assumption_model}
\begin{assumption_model}
We approximate the size of ribosomes to be such that they only occupy a single codon (or RBS) along a transcript and neighbouring codons (and RBSes) can be occupied by separate ribosomes.
\end{assumption_model}
\begin{assumption_model}
No two ribosomes can occupy the same codon or RBS.
\end{assumption_model}
\begin{assumption_model}
Ribosomes move along transcript one codon at a time and cannot move to the next codon if it is occupied by another ribosome.
\end{assumption_model}
\begin{assumption_model}
Ribosomes move from one codon to the next at a fixed and constant rate if the next codon is not occupied.
\end{assumption_model}
\begin{assumption_model}
There is a \textit{large} number of total ribosomes, so $R>>1$
\end{assumption_model}
\begin{assumption_model}
Transitions from one elongation state to the next are single steps and time is modelled discretely with intervals $\delta t$. A maximum of one state transition for each ribosome may occur during this time interval $\delta t$ (i.e. maximum one elongation step per ribosome during the time interval $\delta t$).
\end{assumption_model}
\begin{assumption_model}
All transcripts are identical and so the probability of a ribosome $r$ being in elongation stage $i\in \{0,...,L\}$ at time $t$ is the same for any mRNA:
\[\mathbb{P}(E^{r}_{m,i},t)=\mathbb{P}(E^{r}_{s,i},t)\hspace{1cm}\forall \hspace{0.2cm}m,s \in \{1,...,M\}\]
\end{assumption_model}
\begin{assumption_model}
All ribosomes are identical and so have the same probability distribution:
\[\mathbb{P}(E^{r}_{m,i},t)=\mathbb{P}(E^{q}_{m,i},t) \hspace{1cm}\forall \hspace{0.2cm} r,q \in \{1,...,R\}\]
\end{assumption_model}
\begin{assumption_model}
The event of a new transcript creation is defined as the moment an mRNA is finished being transcribed. Ribosomes move along mRNA closely following the RNA polymerase as it transcribes, and are already moving along the mRNA before it has finished being fully transcribed \cite{proshkincooperation2010}. Therefore upon the creation of a new mRNA we can approximate that it is already covered in ribosomes and we make a steady-state approximation over these equations.
\end{assumption_model}
\begin{assumption_model}
The process is Markovian and at any point in time the position of one ribosome is independent of the positions of others at that point in time.
\end{assumption_model}

While we acknowledge some of these assumptions do not accurately represent the reality of the complex biological process of translation, we make them in order to simplify the model in a way that we do not anticipate will affect the core behaviours of the translation dynamics. For example, we know that a ribosome occupies space that covers more than one codon at a time, however by approximating it as occupying the space of only one codon we simplify the model significantly. What we lose in the accuracy of this specific detail we more than make up for in the increased ease with which we can work with the model.

\begin{figure}
\begin{center}
\includegraphics[width=0.98\textwidth]{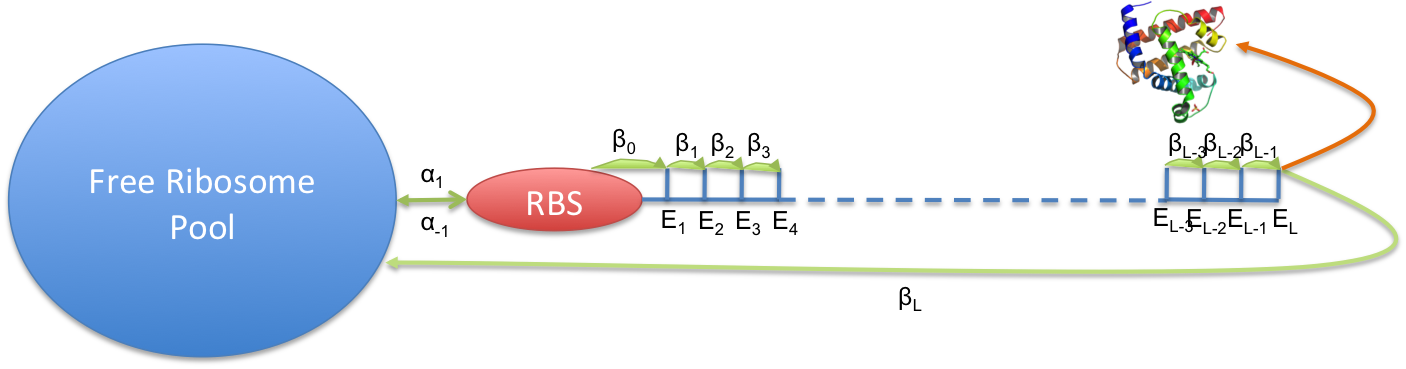}
\end{center}
\caption[Ribosomal movement]{A representation of the flow of ribosomes in translation (green arrows). The blue scaffold represents the mRNA with $E_i$s being codons along it, the green arrows represent the direction of movement of the ribosome with $\alpha_i$s and $\beta_i$s being the rates and the orange arrow represents the production of a protein.}
\label{fig:ribosomes}
\end{figure}

Figure \ref{fig:ribosomes} shows a cartoon schematic of the process we are modelling.

\subsubsection{Events}
\begin{enumerate}
\item $E^{r}_{m,i}$ is the event of ribosome $r$ being on transcript $m$ in elongation state $i$ (i.e. at the $i$\textsuperscript{th} codon) for $i \in \{1,...,L\}$.
\item $\lnot E^{r}_{m,i}$ is the event of ribosome $r$ not being on transcript $m$ in elongation state $i$ (i.e. at the $i$\textsuperscript{th} codon) for $i \in \{1,...,L\}$.
\item $E^{r}_{m,0}$ is the event of ribosome $r$ being on the RBS of transcript $m$.
\item $Rib^{r}$ is the event of ribosome $r$ not being on any transcript (i.e. in the free ribosome pool).
\end{enumerate}

For any ribosome `r' from a pool of R ribosomes, if it is freely available ($Rib^{r}$) it can bind to the RBS of mRNA `m' ($E^{r}_{m,0}$) and from this state it can either unbind and join the free ribosome pool again, or translation can be initiated and it moves into the initial state of elongation ($E^{r}_{m,1}$). From the initial state of elongation $E^{r}_{m,1}$ the only path the ribosome can take is to go from the i\textsuperscript{th} stage of elongation ($E^{r}_{m,i}$) to the i+1\textsuperscript{th} stage of elongation ($E^{r}_{m,i+1}$) until it reaches the final elongation stage which, without loss of generality, can be the L\textsuperscript{th} stage ($E^{r}_{m,L}$). From this, translation finishes, a full protein is produced and the ribosome returns to the free ribosome pool.

$\mathbb{P}({E^{r}_{m,i},t})$ is the probability that event $E^{r}_{m,i}$ occurs at time $t$. For the random walk we consider a discrete time distribution with steps of length $\delta t$ and define the following `rates':

\begin{defi}\label{def1}
We define the `unblocked' rates:
\begin{eqnarray*}
\alpha^+&=&\lim_{\delta t \to 0} \frac{\mathbb{P}\big(E^{r}_{m,0},t+\delta t|Rib^{r},t\cap(\bigcap_{q\neq r}\lnot E^{q}_{m,0},t+\delta t)\big)}{\delta t}\\
\alpha^-&=&\lim_{\delta t \to 0}\frac{\mathbb{P}(Rib^{r},t+\delta t|E^{r}_{m,0},t)}{\delta t}\\
\beta_i&=&\lim_{\delta t \to 0}\frac{\mathbb{P}\big(E^{r}_{m,i+1},t+\delta t|E^{r}_{m,i},t\cap(\bigcap_{q\neq r}\lnot E^{q}_{m,i+1},t+\delta t )\big)}{\delta t} \hspace{0.5cm} \text{ for } i\in \{0,...,L-1\}\\
\beta_L&=&\lim_{\delta t \to 0}\frac{\mathbb{P}(Rib^{r},t+\delta t|E^{r}_{m,L},t)}{\delta t}
\end{eqnarray*} 
\end{defi}

where $\alpha^+$ is the binding rate of a ribosome to an RBS, $\alpha^-$ is the unbinding rate of a ribosome from an RBS and the $\beta_i$ values are the rates of elongation at which a ribosome moves to the next codon (if it is not blocked).

\subsubsection{Proposed translation model}
The system can be displayed mathematically as:
\begin{subequations}
\begin{eqnarray}
\mathbb{P}(Rib^{r},t+\delta t) & = & \mathbb{P}(Rib^{r},t+\delta t|Rib^{r},t)\mathbb{P}(Rib^{r},t) \nonumber \\
& & { } + \sum^M_{s=1}\mathbb{P}(Rib^{r},t+\delta t|E^{r}_{s,0},t)\mathbb{P}(E^{r}_{s,0},t) \nonumber \\
& & {} + \sum^M_{s=1}\mathbb{P}(Rib^{r},t+\delta t|E^{r}_{s,L},t)\mathbb{P}(E^{r}_{s,L},t)\allowdisplaybreaks\\
& & \nonumber \\
\mathbb{P}(E^{r}_{m,0},t+\delta t) & = & \mathbb{P}(E^{r}_{m,0},t+\delta t|E^{r}_{m,0},t)\mathbb{P}(E^{r}_{m,0},t) \nonumber \\
& & { } + \mathbb{P}(E^{r}_{m,0},t+\delta t|Rib^{r},t)\mathbb{P}(Rib^{r},t)\allowdisplaybreaks\\
& & \nonumber \\
\mathbb{P}(E^{r}_{m,1},t+\delta t) & = & \mathbb{P}(E^{r}_{m,1},t+\delta t|E^{r}_{m,1},t)\mathbb{P}(E^{r}_{m,1},t) \nonumber \\
& & { } + \mathbb{P}(E^{r}_{m,1},t+\delta t|E^{r}_{m,0},t)\mathbb{P}(E^{r}_{m,0},t)\allowdisplaybreaks\\
& & \vdots \nonumber \\
\mathbb{P}(E^{r}_{m,i},t+\delta t) & = & \mathbb{P}(E^{r}_{m,i},t+\delta t|E^{r}_{m,i},t)\mathbb{P}(E^{r}_{m,i},t)  \nonumber \\
& & { } + \mathbb{P}(E^{r}_{m,i},t+\delta t|E^{r}_{m,i-1},t)\mathbb{P}(E^{r}_{m,i-1},t) \hspace{1cm} \forall i \in \{2,...,L-1\}\allowdisplaybreaks\\
& & \vdots \nonumber \\
\mathbb{P}(E^{r}_{m,L},t+\delta t) & = & \mathbb{P}(E^{r}_{m,L},t+\delta t|E^{r}_{m,L},t)\mathbb{P}(E^{r}_{m,L},t) \nonumber \\
& & { } + \mathbb{P}(E^{r}_{m,L},t+\delta t|E^{r}_{m,L-1},t)\mathbb{P}(E^{r}_{m,L-1},t)
\end{eqnarray}
\end{subequations}
We next rewrite the probability of a ribosome staying in the same state as being equal to 1 minus the probability of it moving out of that state:
\begin{subequations}
\begin{eqnarray}
\mathbb{P}(Rib^{r},t+\delta t) & = & \left(1-\sum^M_{s=1}\mathbb{P}(E^{r}_{s,0},t+\delta t|Rib^{r},t)\right)\mathbb{P}(Rib^{r},t) \nonumber \\
& & { } + \sum^M_{s=1}\mathbb{P}(Rib^{r},t+\delta t|E^{r}_{s,0},t)\mathbb{P}(E^{r}_{s,0},t) \nonumber \\
& & {} + \sum^M_{s=1}\mathbb{P}(Rib^{r},t+\delta t|E^{r}_{s,L},t)\mathbb{P}(E^{r}_{s,L},t)\allowdisplaybreaks\\
& & \nonumber \\
\mathbb{P}(E^{r}_{m,0},t+\delta t) & = & \left(1-\mathbb{P}(Rib^{r},t+\delta t|E^{r}_{m,0},t)-\mathbb{P}(E^{r}_{m,1},t+\delta t|E^{r}_{m,0},t)\right)\mathbb{P}(E^{r}_{m,0},t) \nonumber \\
& & { } + \mathbb{P}(E^{r}_{m,0},t+\delta t|Rib^{r},t)\mathbb{P}(Rib^{r},t)\allowdisplaybreaks\\
& & \nonumber \\
\mathbb{P}(E^{r}_{m,1},t+\delta t) & = & \left(1 - \mathbb{P}(E^{r}_{m,2},t+\delta t|E^{r}_{m,1},t)\right)\mathbb{P}(E^{r}_{m,1},t) \nonumber \\
& & { } + \mathbb{P}(E^{r}_{m,1},t+\delta t|E^{r}_{m,0},t)\mathbb{P}(E^{r}_{m,0},t)\allowdisplaybreaks\\
& & \vdots \nonumber \\
\mathbb{P}(E^{r}_{m,i},t+\delta t) & = & \left(1 - \mathbb{P}(E^{r}_{m,i+1},t+\delta t|E^{r}_{m,i},t)\right)\mathbb{P}(E^{r}_{m,i},t) \nonumber \\
& & { } + \mathbb{P}(E^{r}_{m,i},t+\delta t|E^{r}_{m,i-1},t)\mathbb{P}(E^{r}_{m,i-1},t) \hspace{1cm} \forall i \in \{2,...,L-1\}\allowdisplaybreaks\\
& & \vdots \nonumber \\
\mathbb{P}(E^{r}_{m,L},t+\delta t) & = & \left(1-\mathbb{P}(Rib^{r},t+\delta t|E^{r}_{m,L},t)\right)\mathbb{P}(E^{r}_{m,L},t) \nonumber \\
& & { } + \mathbb{P}(E^{r}_{m,L},t+\delta t|E^{r}_{m,L-1},t)\mathbb{P}(E^{r}_{m,L-1},t)
\end{eqnarray}
\end{subequations}
Rearranging gives:
\begin{subequations}
\begin{eqnarray}
\mathbb{P}(Rib^{r},t+\delta t) - \mathbb{P}(Rib^{r},t)& = & -\sum^M_{s=1}\mathbb{P}(E^{r}_{s,0},t+\delta t|Rib^{r},t)\mathbb{P}(Rib^{r},t) \nonumber \\
& & { } + \sum^M_{s=1}\mathbb{P}(Rib^{r},t+\delta t|E^{r}_{s,0},t)\mathbb{P}(E^{r}_{s,0},t) \nonumber \\
& & {} + \sum^M_{s=1}\mathbb{P}(Rib^{r},t+\delta t|E^{r}_{s,L},t)\mathbb{P}(E^{r}_{s,L},t)\allowdisplaybreaks\\
& & \nonumber \\
\mathbb{P}(E^{r}_{m,0},t+\delta t) - \mathbb{P}(E^{r}_{m,0},t)& = & -\Big(\mathbb{P}(Rib^{r},t+\delta t|E^{r}_{m,0},t)-\mathbb{P}(E^{r}_{m,1},t+\delta t|E^{r}_{m,0},t)\Big)\mathbb{P}(E^{r}_{m,0},t) \nonumber \\
& & { } + \mathbb{P}(E^{r}_{m,0},t+\delta t|Rib^{r},t)\mathbb{P}(Rib^{r},t)\allowdisplaybreaks\\
& & \nonumber \\
\mathbb{P}(E^{r}_{m,1},t+\delta t) - \mathbb{P}(E^{r}_{m,1},t)& = &  - \mathbb{P}(E^{r}_{m,2},t+\delta t|E^{r}_{m,1},t)\mathbb{P}(E^{r}_{m,1},t) \nonumber \\
& & { } + \mathbb{P}(E^{r}_{m,1},t+\delta t|E^{r}_{m,0},t)\mathbb{P}(E^{r}_{m,0},t)\allowdisplaybreaks\\
& & \vdots \nonumber \\
\mathbb{P}(E^{r}_{m,i},t+\delta t) - \mathbb{P}(E^{r}_{m,i},t)& = &  - \mathbb{P}(E^{r}_{m,i+1},t+\delta t|E^{r}_{m,i},t)\mathbb{P}(E^{r}_{m,i},t)\mbox{ {         }   for } i \in \{2,...,L-1\}  \nonumber \\
& & { } + \mathbb{P}(E^{r}_{m,i},t+\delta t|E^{r}_{m,i-1},t)\mathbb{P}(E^{r}_{m,i-1},t)\allowdisplaybreaks\\
& & \vdots \nonumber \\
\mathbb{P}(E^{r}_{m,L},t+\delta t) - \mathbb{P}(E^{r}_{m,L},t)& = & -\mathbb{P}(Rib^{r},t+\delta t|E^{r}_{m,L},t)\mathbb{P}(E^{r}_{m,L},t) \nonumber \\
& & { } + \mathbb{P}(E^{r}_{m,L},t+\delta t|E^{r}_{m,L-1},t)\mathbb{P}(E^{r}_{m,L-1},t)
\end{eqnarray}
\label{eqns3}
\end{subequations}

For all events $E^{r}_{m,i}$ at a time $t+\delta t$ we have that the probability $\mathbb{P}(E^{r}_{m,i},t+\delta t|X)$ for any event $X$ can be split into two subsets, one where there is a ribosome in elongation state $i$ on mRNA $m$ at time $t$ and one where there is not:
\begin{eqnarray}
\mathbb{P}(E^{r}_{m,i},t|X)\mathbb{P}(X) & = & \mathbb{P}(E^{r}_{m,i},t|X\cap(\bigcup_{q\neq r}E^{q}_{m,i},t) )\mathbb{P}(X\cap(\bigcup_{q\neq r}E^{q}_{m,i},t) ) \nonumber\\
& & + \mathbb{P}(E^{r}_{m,i},t|X\cap(\bigcap_{q\neq r}\lnot E^{q}_{m,i},t) )\mathbb{P}(X\cap(\bigcap_{q\neq r}\lnot E^{q}_{m,i},t))
\end{eqnarray}
Since the probability of two ribosomes being in the same state on the same mRNA is zero we must have that:
\begin{eqnarray}
\mathbb{P}(E^{r}_{m,i},t|X)\mathbb{P}(X) & = &  \mathbb{P}(E^{r}_{m,i},t|X\cap(\bigcap_{q\neq r}\lnot E^{q}_{m,i},t) )\mathbb{P}(X\cap(\bigcap_{q\neq r}\lnot E^{q}_{m,i},t ))
\end{eqnarray}
which can be rewritten as:
\begin{eqnarray}
\mathbb{P}(E^{r}_{m,i},t|X)\mathbb{P}(X) & = &  \mathbb{P}(E^{r}_{m,i},t|X\cap(\bigcap_{q\neq r}\lnot E^{q}_{m,i},t ) )\mathbb{P}(\bigcap_{q\neq r}\lnot E^{q}_{m,i},t |X)\mathbb{P}(X)
\end{eqnarray}
Due to mutual exclusivity, we know that the probability of no other ribosomes being there is equal to 1 minus the sum of the probabilities of each other ribosome being there:
\begin{eqnarray}
\mathbb{P}(E^{r}_{m,i},t|X)\mathbb{P}(X) & = &  \mathbb{P}(E^{r}_{m,i},t|X\cap(\bigcap_{q\neq r}\lnot E^{q}_{m,i},t ) )(1-\sum_{q\neq r}\mathbb{P}(E^{q}_{m,i},t|X))\mathbb{P}(X)
\end{eqnarray}
Combining this with equations \eqref{eqns3} gives:
\begin{subequations}
\begin{eqnarray}
\mathbb{P}(Rib^{r},t+\delta t) - \mathbb{P}(Rib^{r},t)& = & -\sum^M_{s=1}\Bigg(\mathbb{P}\big(E^{r}_{s,0},t+\delta t|Rib^{r},t\cap(\bigcap_{q\neq r}\lnot E^{q}_{s,0},t+\delta t)\big)\nonumber \allowdisplaybreaks\\
& & \hspace{1cm} \cdot\Big(1-\sum_{q\neq r}\mathbb{P}(E^{q}_{s,0},t+\delta t|Rib^{r},t)\Big)\mathbb{P}(Rib^{r},t) \Bigg)\nonumber\allowdisplaybreaks \\
& & { } + \sum^M_{s=1}\mathbb{P}(Rib^{r},t+\delta t|E^{r}_{s,0},t)\mathbb{P}(E^{r}_{s,0},t) \nonumber \allowdisplaybreaks\\
& & {} + \sum^M_{s=1}\mathbb{P}(Rib^{r},t+\delta t|E^{r}_{s,L},t)\mathbb{P}(E^{r}_{s,L},t)\allowdisplaybreaks\\
& & \nonumber\allowdisplaybreaks \\
\mathbb{P}(E^{r}_{m,0},t+\delta t) - \mathbb{P}(E^{r}_{m,0},t)& = & -\mathbb{P}(Rib^{r},t+\delta t|E^{r}_{m,0},t)\mathbb{P}(E^{r}_{m,0},t)\nonumber \allowdisplaybreaks\\
& & {}-\bigg(\mathbb{P}\big(E^{r}_{m,1},t+\delta t|E^{r}_{m,0},t\cap(\bigcap_{q\neq r}\lnot E^{q}_{m,1},t+\delta t )\big) \nonumber\allowdisplaybreaks \\
& & \hspace{1cm} \cdot\Big(1-\sum_{q\neq r}\mathbb{P}(E^{q}_{m,1},t |E^{r}_{m,0},t)\Big)\mathbb{P}(E^{r}_{m,0},t) \bigg)\nonumber \allowdisplaybreaks\\
& & { } + \bigg(\mathbb{P}\big(E^{r}_{m,0},t+\delta t|Rib^{r},t\cap(\bigcap_{q\neq r}\lnot E^{q}_{m,0},t+\delta t )\big)\nonumber\allowdisplaybreaks \\
& & \hspace{1cm} \cdot\Big(1-\sum_{q\neq r}\mathbb{P}(E^{q}_{m,0},t |Rib^{r},t)\Big)\mathbb{P}(Rib^{r},t) \bigg)\allowdisplaybreaks\\
& & \nonumber \allowdisplaybreaks\\
\mathbb{P}(E^{r}_{m,1},t+\delta t) - \mathbb{P}(E^{r}_{m,1},t)& = & {}-\bigg(\mathbb{P}\big(E^{r}_{m,2},t+\delta t|E^{r}_{m,1},t\cap(\bigcap_{q\neq r}\lnot E^{q}_{m,2},t+\delta t)\big) \nonumber\allowdisplaybreaks \\
& & \hspace{1cm} \cdot\Big(1-\sum_{q\neq r}\mathbb{P}(E^{q}_{m,2},t |E^{r}_{m,1},t)\Big)\mathbb{P}(E^{r}_{m,1},t) \bigg)\nonumber \allowdisplaybreaks\\
& & { } + \bigg(\mathbb{P}\big(E^{r}_{m,1},t+\delta t|E^{r}_{m,0},t\cap(\bigcap_{q\neq r}\lnot E^{q}_{m,1},t+\delta t)\big) \nonumber\allowdisplaybreaks \\
& & \hspace{1cm} \cdot\Big(1-\sum_{q\neq r}\mathbb{P}(E^{q}_{m,1},t |E^{r}_{m,0},t)\Big)\mathbb{P}(E^{r}_{m,0},t) \bigg) \allowdisplaybreaks\\
& & \vdots \nonumber\allowdisplaybreaks \\
\mathbb{P}(E^{r}_{m,i},t+\delta t) - \mathbb{P}(E^{r}_{m,i},t)& = &  - \bigg(\mathbb{P}\big(E^{r}_{m,i+1},t+\delta t|E^{r}_{m,i},t\cap(\bigcap_{q\neq r}\lnot E^{q}_{m,i+1},t+\delta t )\big) \nonumber\allowdisplaybreaks \\
& & \hspace{1cm} \cdot\Big(1-\sum_{q\neq r}\mathbb{P}(E^{q}_{m,i+1},t |E^{r}_{m,i},t)\Big)\mathbb{P}(E^{r}_{m,i},t) \bigg)\nonumber\allowdisplaybreaks \\
& & { } + \bigg(\mathbb{P}\big(E^{r}_{m,i},t+\delta t|E^{r}_{m,i-1},t\cap(\bigcap_{q\neq r}\lnot E^{q}_{m,i},t+\delta t )\big) \nonumber \allowdisplaybreaks\\
& & \hspace{1cm} \cdot\Big(1-\sum_{q\neq r}\mathbb{P}(E^{q}_{m,i},t |E^{r}_{m,i-1},t)\Big)\mathbb{P}(E^{r}_{m,i-1},t) \bigg) \hspace{1cm} \\
& & \forall i \in \{2,...,L-1\} \allowdisplaybreaks\\
& & \vdots \nonumber \allowdisplaybreaks\\
\mathbb{P}(E^{r}_{m,L},t+\delta t) - \mathbb{P}(E^{r}_{m,L},t)& = & -\mathbb{P}(Rib^{r},t+\delta t|E^{r}_{m,L},t)\mathbb{P}(E^{r}_{m,L},t) \nonumber \\
& & { } + \bigg(\mathbb{P}\big(E^{r}_{m,L},t+\delta t|E^{r}_{m,L-1},t\cap(\bigcap_{q\neq r}\lnot E^{q}_{m,L},t+\delta t)\big) \nonumber \allowdisplaybreaks\\
& & \hspace{1cm} \cdot\Big(1-\sum_{q\neq r}\mathbb{P}(E^{q}_{m,L},t |E^{r}_{m,L-1},t)\Big)\mathbb{P}(E^{r}_{m,L-1},t) \bigg) 
\end{eqnarray}
\end{subequations}
Dividing both sides by $\delta t$ and taking $\lim_{\delta t \to 0}$ as well as taking the definitions of the rates as mentioned in Definition \ref{def1} gives:
\begin{subequations}
\begin{eqnarray}
\frac{d\mathbb{P}(Rib^{r},t)}{dt} & = &  \sum^M_{s=1} \alpha^{-}\mathbb{P}(E^{r}_{s,0},t) \nonumber \\
& & { } -  \sum^M_{s=1} \alpha^{+}\mathbb{P}(Rib^{r},t)(1 - \sum_{q\neq r} \mathbb{P}(E^{q}_{s,0},t)) \nonumber \\
& & { } +  \sum^M_{s=1} \beta_{L}\mathbb{P}(E^{r}_{s,L},t) \allowdisplaybreaks\\
\frac{d\mathbb{P}(E^{r}_{m,0},t)}{dt} & = & -\alpha^{-}\mathbb{P}(E^{r}_{m,0},t) \nonumber \\
& & { } +\alpha^{+}\mathbb{P}(Rib^{r},t)(1 - \sum_{q\neq r} \mathbb{P}(E^{q}_{m,0},t)) \nonumber \\
& & { } - \beta_0\mathbb{P}(E^{r}_{m,0},t))(1-\sum_{q\neq r}\mathbb{P}(E^{q}_{m,1},t))\allowdisplaybreaks\\
\frac{d\mathbb{P}(E^{r}_{m,1},t)}{dt} & = & \beta_0\mathbb{P}(E^{r}_{m,0},t)(1-\sum_{q\neq r} \mathbb{P}(E^{q}_{m,1},t)) \nonumber \\
& & { } - \beta_1\mathbb{P}(E^{r}_{m,1},t)(1-\sum_{q\neq r} \mathbb{P}(E^{q}_{m,2},t))\allowdisplaybreaks\\
& & \vdots \nonumber \\
\frac{d\mathbb{P}(E^{r}_{m,i},t)}{dt} & = & \beta_{i-1}\mathbb{P}(E^{r}_{m,i-1},t)(1-\sum_{q\neq r} \mathbb{P}(E^{q}_{m,i},t)) \nonumber \\
& & { } - \beta_i\mathbb{P}(E^{r}_{m,i},t)(1-\sum_{q\neq r} \mathbb{P}(E^{q}_{m,i+1},t))\allowdisplaybreaks\\
& & \vdots \nonumber \\
\frac{d\mathbb{P}(E^{r}_{m,L},t)}{dt} & = & \beta_{L-1}\mathbb{P}(E^{r}_{m,L-1},t)(1-\sum_{q\neq r} \mathbb{P}(E^{q}_{m,L},t)) \nonumber \\
& & { } - \beta_L\mathbb{P}(E^{r}_{m,L},t)
\end{eqnarray}
\label{probodes}
\end{subequations}
Using assumptions (4) and (5) we can take the sums and along with using assumption (6) and saying $R>>1\implies R-1\simeq R$ (the minimal number of ribosomes we have modelled this system with is 1000, and often higher meaning this assumption holds) we get:
\begin{subequations}
\begin{eqnarray}
\frac{d\mathbb{P}(Rib^{r},t)}{dt} & = &  M\cdot \alpha^{-}\mathbb{P}(E^{r}_{m,0},t) \nonumber \\
& & { } -  M\cdot \alpha^{+}\mathbb{P}(Rib^{r},t)(1 - R\cdot \mathbb{P}(E^{r}_{m,0},t)) \nonumber \\
& & { } +  M\cdot \beta_{L}\mathbb{P}(E^{r}_{m,L},t) \allowdisplaybreaks\\
\frac{d\mathbb{P}(E^{r}_{m,0},t)}{dt} & = & -\alpha^{-}\mathbb{P}(E^{r}_{m,0},t) \nonumber \\
& & { } +\alpha^{+}\mathbb{P}(Rib^{r},t)(1 - R\cdot \mathbb{P}(E^{r}_{m,0},t)) \nonumber \\
& & { } - \beta_0\mathbb{P}(E^{r}_{m,0},t))(1-R\cdot \mathbb{P}(E^{r}_{m,1},t))\allowdisplaybreaks\\
\frac{d\mathbb{P}(E^{r}_{m,1},t)}{dt} & = & \beta_0\mathbb{P}(E^{r}_{m,0},t)(1-R\cdot \mathbb{P}(E^{r}_{m,1},t)) \nonumber \\
& & { } - \beta_1\mathbb{P}(E^{r}_{m,1},t)(1- R\cdot \mathbb{P}(E^{r}_{m,2},t))\allowdisplaybreaks\\
& & \vdots \nonumber \\
\frac{d\mathbb{P}(E^{r}_{m,i},t)}{dt} & = & \beta_{i-1}\mathbb{P}(E^{r}_{m,i-1},t)(1-R\cdot \mathbb{P}(E^{r}_{m,i},t)) \nonumber \\
& & { } - \beta_i\mathbb{P}(E^{r}_{m,i},t)(1-R\cdot \mathbb{P}(E^{r}_{m,i+1},t))\allowdisplaybreaks\\
& & \vdots \nonumber \\
\frac{d\mathbb{P}(E^{r}_{m,L},t)}{dt} & = & \beta_{L-1}\mathbb{P}(E^{r}_{m,L-1},t)(1-R\cdot \mathbb{P}(E^{r}_{m,L},t)) \nonumber \\
& & { } - \beta_L\mathbb{P}(E^{r}_{m,L},t)
\end{eqnarray}
\label{probodes}
\end{subequations}

A set of random variables $X_{m,i}$ ($i \in \{0...L\}$) is defined as follows:
\[
 X_{m,i}(t) = \left\{ 
  \begin{array}{l l}
    1 & \quad \text{if there is a ribosome present in elongation stage `$i$' on mRNA `$m$' at time `$t$'}\\
    0 & \quad \text{if there is no ribosome present in elongation stage `$i$' on mRNA `$m$' at time `$t$'}\\
  \end{array} \right.
\]
The random variable $F(t)$ represents the number of ribosomes not on a transcript at time $t$.
At any time `$t$', using assumption (17) on independence of ribosome positions, we have: 
\[\mathbb{P}(X_{m,i}(t) = 1)=\sum_{r}^{}\mathbb{P}(E^{r}_{m,i},t)\]
and
\[\mathbb{P}(X_{m,i}(t) = 0)=1 - \sum_{r}^{}\mathbb{P}(E^{r}_{m,i},t)\]

By the definition of expectation, we then have:
\[\mathbb{E}(X_{m,i}(t)) = 1\cdot \mathbb{P}(X_{m,i}(t) = 1) + 0\cdot \mathbb{P}(X_{m,i}(t) = 0)\]
so that,
\[\mathbb{E}(X_{m,i}(t)) = \sum_{r}^{}\mathbb{P}(E^{r}_{m,i},t)\]
Using assumption (1) we get:
\begin{eqnarray}
\mathbb{E}(X_{m,i}(t)) = R\cdot \mathbb{P}(E^{r}_{m,i},t) \label{eqn1}
\end{eqnarray}
where $R$ is the total number of ribosomes. We further define the random variable $X_{i}(t)$ as the sum of random variables $X_{m,i}(t)$ across all mRNA, i.e. the total number of ribosomes in position $i$ across all transcripts:
\[X_{i}(t)=\sum_{m}^{}X_{m,i}(t)\]
which, taking expectations, gives
\[\mathbb{E}(X_{i}(t))=\sum_{m}\mathbb{E}(X_{m,i}(t))\]
Combining with equation \eqref{eqn1} gives
\begin{eqnarray}
\mathbb{E}(X_{i}(t))=\sum_{m}R\cdot \mathbb{P}(E^{r}_{m,i},t)
\end{eqnarray}
Now, using assumption (2), we have:
\begin{eqnarray}
\mathbb{E}(X_{i}(t)) = MR\cdot \mathbb{P}(E^{r}_{m,i},t) \
\end{eqnarray}

We define the variable $Y_i(t)$ to be the expectation of the random variable $X_i(t)$
\begin{eqnarray}
Y_i(t) = \mathbb{E}(X_{i}(t)) 
\end{eqnarray}
Therefore,
\begin{eqnarray}
Y_i (t)= MR\cdot \mathbb{P}(E^{r}_{m,i},t)  & \forall i \in \{0,...,L\}
\label{Xi}
\end{eqnarray}

We now investigate the variance of $X_i(t)$. The variance of each $X_{m,i}(t)$ is equal to:

\begin{eqnarray}
Var(X_{m,i}(t)) & = & \mathbb{E}\left(X_{m,i}(t)^2\right) - \mathbb{E}\left(X_{m,i}(t)\right)^2
\end{eqnarray}

However, since $X_{m,i}(t)$ can only take the values $0$ or $1$, it must be true that $X_{m,i}(t)^2 = X_{m,i}(t)$ and so, if we let $\mu = \mathbb{E}(X_{m,i}(t))$, we obtain:

\begin{eqnarray}
Var(X_{m,i}(t)) & = & \mu - \mu^2
\end{eqnarray}

$X_i(t)$ is a random variable that represents the sum of a population of  independent, identically distributed (IID) random variables $X_{m,i}(t)$. From the variance of a population of independent, identically distributed random variables we get:

\begin{eqnarray}
Var(X_{i}(t)) & = &  \frac{\mu - \mu^2}{M}
\end{eqnarray}

This gives us an estimate of the cell to cell variance we would expect from this model. This indicates that the behaviour of the circuit becomes less noisy as the number of transcripts increases since the variance per cell is inversely proportional to the number of transcripts. This is an interesting result and suggests that a stronger promoter would cause lower cell to cell variation.
\\

\begin{rem}
There are many other factors not included in this model which can cause additional cell to cell variations as well as population level variations in gene circuit output.

It is tempting to include a `deterministic' variance to estimate what the cell to cell variation would be. However, simulations from our model are performed to represent populations of identical cells. Therefore the cell to cell variance captured by our model cannot be compared to the population variance typically observed in experimental population measurements of \emph{in vivo} gene expression. This is because cell populations contain very large numbers of cells in which any variation as predicted by our model would be silenced. The variation observed experimentally is typically due to additional factors not included in our model.
\end{rem}

The random variable $F$, which represents the number of free ribosomes, can be calculated as the total number of ribosomes minus the expected total number of ribosomes on transcripts:
\begin{eqnarray}
F (t) = R - \sum_{i}X_i (t)
\label{F}
\end{eqnarray}
Letting $G(t)$ be the expectation of the random variable $F(t)$
\begin{eqnarray}
G(t) = \mathbb{E}(F(t)),
\end{eqnarray}
combining \eqref{Xi} and \eqref{F} with \eqref{probodes}, and dropping the $(t)$ from the notation by letting $X_i=X_i(t)$ and $F=F(t)$ we are left with:

\begin{subequations}
\begin{eqnarray}
\frac{dG}{dt} & = & -M\alpha^+G(1-Y_{0}/M) + \alpha^-Y_{0} + \beta_LY_{L}\allowdisplaybreaks\\
\frac{dY_{0}}{dt} & = & M\alpha^+G(1-Y_{0}/M) - \alpha^-Y_{0} - \beta_0Y_{0}(1-Y_{1}/M)\allowdisplaybreaks\\
\frac{dY_{1}}{dt} & = & \beta_0Y_{0}(1-Y_{1}/M) - \beta_1Y_{1}(1-Y_{2}/M)\allowdisplaybreaks\\
& & \vdots \nonumber \\
\frac{dY_{1}}{dt} & = & \beta_{i-1}Y_{i-1}(1-Y_{i}/M) - \beta_iY_{i}(1-Y_{i+1}/M)\allowdisplaybreaks\\
& & \vdots \nonumber \\
\frac{dY_{L}}{dt} & = & \beta_{L-1}Y_{L-1}(1-Y_{L}/M) - \beta_LY_{L}
\end{eqnarray}
\end{subequations}

The steady state equations, obtained assuming that the system is in exponential growth and that each transcript has a steady state distribution of ribosomes on it (Assumption 15), are then given by:

\begin{eqnarray}
M\alpha^+G(1-Y_{0}/M) & = & \alpha^-Y_{0} + \beta_LY_{L}\allowdisplaybreaks\\
\beta_0Y_{0}(1-Y_{1}/M) & = & \beta_1Y_{1}(1-Y_{2}/M)\nonumber\\
& = & \vdots\nonumber\\
& = & \beta_{i-1}Y_{i-1}(1-Y_{i}/M) \nonumber\\
& = & \vdots\nonumber\\
& = & \beta_{L-1}Y_{L-1}(1-Y_{L}/M) \nonumber\\
& = &\beta_LY_{L}
\label{finalmodel}
\end{eqnarray}

%
%
%

\subsection{Solving the steady-state equations}

\label{sec:translation_model}

Rearranging Equation \eqref{finalmodel} and letting

\[Rib = G\]
\[L = Y_0\]
\[F_i = X_i\]
we can define functions $f_{Rib}$, $f_L$ and $f_k \hspace{01cm} \forall k \in \{1,...,m-1\}$:

\begin{subequations}
\begin{eqnarray}
Rib & = & f_{Rib}(L,F_1) \allowdisplaybreaks\\
L & = & f_L(F_1,F_2) \allowdisplaybreaks\\
F_k & = & f_k(F_{k+1},F_{k+2})\hspace{1cm} \forall k \in \{1,...,m-2\} \allowdisplaybreaks\\
F_{m-1} & = & f_{m-1}(F_m) \allowdisplaybreaks
\end{eqnarray}
\end{subequations}

These functions can be rewritten as:

\begin{subequations}
\begin{eqnarray}
Rib & = & g_{Rib}(F_m) \allowdisplaybreaks\\
L & = & g_L(F_m) \allowdisplaybreaks\\
F_k & = & g_k(F_m)\hspace{1cm} \forall k \in \{1,...,m-2\} \allowdisplaybreaks\\
F_{m-1} & = & g_{m-1}(F_m) \allowdisplaybreaks
\end{eqnarray}
\label{eqn:uniqueness}
\end{subequations}

\subsection{Proving uniqueness of the steady-state solution}

In order to prove that there is a unique solution to the set of equations \eqref{eqn:uniqueness} we must ensure that the total number of ribosomes can be expressed as a strictly monotonically increasing function of the number of ribosomes in the free pool. Then using the inverse function theorem, we can prove uniqueness of the solution to the set of equations \eqref{eqn:uniqueness}. 

Let us first prove strict monotonicity of $F_{m-1}$ as a function of $F_m$:

\begin{subequations}
\begin{eqnarray}
F_{m-1} & = & g_{m-1}(F_m)  \allowdisplaybreaks\\
& = & \frac{\beta_{m}F_{m}}{\beta_{m-1}(R^T-F_{m})}\allowdisplaybreaks
\end{eqnarray}
\end{subequations}
\begin{subequations}
\begin{eqnarray}
\frac{dF_{m-1}}{dF_m} & = & \frac{(\beta_m/\beta_{m-1})R^T}{(R^T-F_{m})^2}\\
& & >0 
\end{eqnarray}
\end{subequations}
So $F_{m-1}$ is a strictly monotonically increasing function of $F_m$. For $F_{m-2}$:
\begin{subequations}
\begin{eqnarray}
F_{m-2} & = & g_{m-2}(F_m)  \allowdisplaybreaks\\
& = & \frac{\beta_{m}F_{m}}{\beta_{m-2}(R^T-F_{m-1})}\allowdisplaybreaks\\
& = & \frac{\beta_{m}F_{m}}{\beta_{m-2}(R^T-\frac{\beta_{m}F_{m}}{\beta_{m-1}(R^T-F_{m})})}\allowdisplaybreaks
\end{eqnarray}
\end{subequations}
\begin{subequations}
\begin{eqnarray}
\frac{dF_{m-2}}{dF_m} & = & (\beta_m/\beta_{m-2})\frac{R^T-(F_{m-1}-F_m\frac{dF_{m-1}}{dF_m})}{(R^T-F_{m-1})^2}   
\end{eqnarray}
\end{subequations}
But we have that:
\begin{subequations}
\begin{eqnarray}
F_{m-1}-F_m\frac{dF_{m-1}}{dF_m} & = & \frac{\beta_{m}}{\beta_{m-2}}\frac{F_{m}}{(R^T-F_{m})} - F_m\frac{R^T}{(R^T-F_{m})^2}\allowdisplaybreaks\\
& = & \frac{\beta_{m}}{\beta_{m-2}}\frac{F_mR^T-F_m^2-F_mR^T}{(R^T-F_{m})^2}\allowdisplaybreaks\\
& = & - \frac{\beta_{m}}{\beta_{m-2}}\frac{F_m^2}{(R^T-F_{m})^2}\allowdisplaybreaks\\
& & \le 0
\end{eqnarray}
\end{subequations}
Therefore
\begin{equation}
\frac{dF_{m-2}}{dF_m} > 0
\end{equation}
and $F_{m-2}$ is, again, a strictly monotonically increasing function of $F_m$. Generally for $F_i$:

Proceeding by induction, we can prove that if $F_{i+2}$ is a strictly monotonically increasing function of $F_m$, then so is $F_i$.

To see this, let's first assume:
\begin{equation}
\frac{dF_{i+2}}{dF_m} > 0
\label{eqn:monotone_assumption}
\end{equation}

We know that:
\begin{subequations}
\begin{eqnarray}
F_{i} & = & g_{i}(F_m)  \allowdisplaybreaks\\
& = & (\beta_m/\beta_i)\frac{F_m}{R^T-F_{i+1}}
\end{eqnarray}
\end{subequations}
Differentiating $F_i$ with respect to $F_m$, we thus have:
\begin{subequations}
\begin{eqnarray}
\frac{dF_{i}}{dF_m} & = & \frac{\beta_m}{\beta_i}\frac{R^T-(F_{i+1}-F_m\frac{dF_{i+1}}{dF_m})}{(R^T-F_{i+1})^2}
\end{eqnarray}
\end{subequations}
Expanding and taking into account Equation \eqref{eqn:monotone_assumption}:
\begin{subequations}
\begin{eqnarray}
F_{i+1}-F_m\frac{dF_{i+1}}{dF_m} & = & \frac{\beta_m}{\beta_{i+2}}(\frac{F_m}{(R^T-F_{i+2})}-F_m\frac{R^T-F_{i+2}+F_m\frac{dF_{i+2}}{dF_m}}{(R^T-F_{i+2})^2})\\
& = & \frac{\beta_m}{\beta_{i+2}}\frac{R^TF_m -F_mF_{i+2}-F_mR^T+F_mF_{i+2}-F_M^2\frac{dF_{i+2}}{dF_m}} {(R^T-F_{i+2})^2}\\
& = & -\frac{\beta_m}{\beta_{i+2}}\frac{F_M^2\frac{dF_{i+2}}{dF_m}} {(R^T-F_{i+2})^2}\\
&  \le & 0
\end{eqnarray}
\end{subequations}
This in turn implies:
\begin{equation}
\frac{dF_{i}}{dF_m} > 0
\end{equation}
Since $\frac{dF_m}{dF_m} > 0$ and $\frac{dF_{m-1}}{dF_m} > 0$ we can use inductive reasoning to get that $
g_{Rib}$, $g_L$ and $g_k$ for $\forall k \in \{1,...,m\}$ are all strictly monotonically increasing functions of $F_m$ and therefore (by the inverse function theorem) have inverse functions. This means we can rewrite all the variables as strictly monotonically increasing functions of $Rib$:
\begin{subequations}
\begin{eqnarray}
Rib & = & h_{Rib}(Rib) \allowdisplaybreaks\\
L & = & h_L(Rib) \allowdisplaybreaks\\
F_k & = & h_k(Rib)\hspace{1cm} \forall k \in \{1,...,m\} \allowdisplaybreaks
\end{eqnarray}
\label{eqn:ribfunctions}
\end{subequations}

Conservation of ribosomes imposes:
\begin{equation}
Rib + h_L(Rib) + \sum_{k=1}^{m}h_k(Rib) = Rib^T
\label{eqn:ribfunction}
\end{equation}

The left-hand-side of this equation is a sum of strictly monotonically increasing functions and therefore is itself a strictly monotonically increasing function of $Rib$ which tends to $+\infty$ as $Rib$  tends to $+\infty$. Therefore, for any $Rib^T$ we have a unique solution for $Rib$ and therefore, from equations \eqref{eqn:uniqueness}, a unique solutions for $L$ and $F_k$, $\forall k \in \{1,...,m\}$.

%
%
%
%
%
%
%

\section{Simulating how a circuit's behaviour impacts on the expression of an unregulated gene}

Equations \eqref{eqn:ribfunctions} and \eqref{eqn:ribfunction} cannot be solved analytically for systems that are large enough to be representative of realistic synthetic circuits. Therefore, the model must be numerically simulated to understand how changing some control points affects the circuit behaviour and free ribosome pool.

A python script was built that allowed this model to be simulated. It consists of two classes, \texttt{Circuit} and \texttt{Cell}. The \texttt{Circuit} class describes individual circuits and allows a user to define the number of transcripts, elongation rates, and binding and unbinding affinities for the RBS. This can be done for any number of circuits. The \texttt{Cell} class allows a user to define a model of a cell including the total number of ribosomes available in the cell as well as which circuit(s) it contains. These classes have attributes and methods that allow simulations of the system to be run. The method \texttt{simulate} on the \texttt{Cell} class allows the simulation of the cell to be run and gives a dictionary output that describes the number of free ribosomes remaining in the cell as well as the distribution of ribosomes for each circuit. This script uses functions provided by the \texttt{scipy} python package, which thus must be installed \emph{a priori.}

\begin{verbatim}
from scipy.optimize import fsolve

###############################
# Ribosomal Competition Model #
###############################

class Cell(object):
    """
    Object respresenting a cell.

    free_ribosomes = integer number of ribosomes available in the cell 
        for synthetic circuits to use.

    circuits = list of synthetic gene circuits in cell.

    """
    def __init__(self, free_ribosomes=1000, circuits=[]):
        self.free_ribosomes = free_ribosomes
        self.circuits = circuits
        self.array = [0] #free ribosomes
        for circuit in circuits:
            self.array += [0]
            
    def conservation(self,p):
        return (sum(p)-self.free_ribosomes,)

    def equation(self,p):
        circuit_lengths = []
        q = [p[0]]
        shift_counter = 1
        for circuit in self.circuits:
            length = circuit.length + 1
            q += [[p[shift_counter:shift_counter+length]]]
            shift_counter += length
        eqns = self.conservation(p)
        for i, circuit in enumerate(self.circuits):
            eqns = eqns + circuit.equation(q,i)
        return eqns

    def simulate(self):
        initial_conditions = [self.free_ribosomes]
        for circuit in self.circuits:
            initial_conditions += circuit.initial_conditions
        solution = fsolve(self.equation, initial_conditions)
        result = {'free_ribosomes': solution[0], 'circuits': []}
        shift_counter = 1
        for circuit in self.circuits:
            length = circuit.length+1
            result['circuits'] += [solution[shift_counter:shift_counter +
                                    length]]
            shift_counter += length
        return result


class Circuit(object):
    """
    Object representing synthetic gene circuits that will be placed into
    a cell.

    total_transcripts = integer number of transcripts for this circuit
        in the cell. Is a function of both copy number and promoter
        strength.

    alpha_plus = rate at which free ribosomes bind to RBS.

    alpha_minus = rate at which ribosomes unbind from RBS.

    betas = list of rates for ribosomes moving along transcript. betas[0]
        represents the rate at which ribosome moves from RBS to initial
        elongation state. betas[i] represents rate at which ribosomes 
        moves from position i to position i+1 if unblocked (or into free 
        ribosome pool for i = length(betas)).

    RBS_strength = single number that replaces alpha_plus, alpha_minus
        and betas[0] if defined.

    """
    def __init__(self, total_transcripts, alpha_plus_scale=0.00001, 
                 alpha_minus_scale=60, length=100, betas=None,
                 RBS_strength=None, speed=20):
        self.total_transcripts = total_transcripts
        if betas:
            self.betas = betas
            self.length = len(betas)-1
        else:
            self.betas = [speed for i in range(length+1)]
            self.length = length
        if RBS_strength:
            self.betas[0] = RBS_strength
        self.alpha_plus = alpha_plus_scale*speed*self.betas[0]
        self.alpha_minus = alpha_minus_scale*speed/self.betas[0]
        self.initial_conditions = [0 for i in range(self.length+1)]]

    def equation(self,q,index):
        """
        Provides ODE equation set for this species solving fsolve
        p is the equation input
        index is index of equation in list for master fsolve
        """
        L = self.length
        T = self.total_transcripts
        a_p = self.alpha_plus
        a_m = self.alpha_minus
        b = self.betas
        x = q[0]
        y = q[index+1][0]
        eqns = ( a_p * x * (T - y[0]) 
                    - a_m * y[0] 
                    - b[0] * y[0] * (1 - y[1]/T) ,)
        for eqn in (( b[i] * y[i] * (1 - y[i+1]/T) 
                        - b[i+1] * y[i+1] * (1 - y[i+2]/T) ,) 
                        for i in range(L-1)):
            eqns = eqns + eqn
        eqns = eqns + (b[L-1] * y[L-1] * (1 - y[L]/T) - b[L] * y[L],)
        return eqns
\end{verbatim}

It is trivial to extend the model described in Section \ref{sec:translation_model} to a system of two (or more) circuits. A simulation was done of a two circuit system in a way where one circuit represented an
unregulated gene that could be easily measured to give information on cell state (called hereafter \textbf{``monitor''}) and the other represented a synthetic gene circuit whose design and  part-composition could be varied (called hereafter \textbf{``circuit''}). Simulation of this model allows predictions to be made about changes in the behaviour of a synthetic circuit when the key control points discussed in Section \ref{sec:control_points} are altered as well as how the expected output from the monitor changes.

\subsection{Parameter and unit checking}

In order to test this model, we start by performing a simulation with realistic values that are observed in the actual system. Using parameters obtained from \href{http://bionumbers.hms.harvard.edu/}{BioNumbers} \cite{bionumbers}, we run a simulation to test whether the output values observed are within realistic bounds. Table \ref{tab:model_validation_parameters} shows the parameters. These roughly represent a medium copy plasmid (25-50 copies per cell) with a medium promoter (2-4 transcripts per promoter in a cell at any time) giving 100 transcripts, 1000 available ribosomes (5\% of total cellular ribosomes at 20,000), a 900 bp CDS (300 amino acids long) that has been codon optimised so elongation rate at each codon is 20 codons per second for all codons.

\begin{table}[H]
\begin{center}
\begin{tabular}{| l | c | l | r | }
\hline
\textbf{Parameter} & \textbf{Model Parameter} & \textbf{Value} & \textbf{Units} \\
\hline
Codon speed (elongation rate) & $\beta_i$ for all $i$ & 20 & ribosomes\textsuperscript{-1} s\textsuperscript{-1} \\
Transcripts & $R^T$ & 100 & mRNA cell\textsuperscript{-1} \\
Available ribosomes & $Rib^T$ & 1000 & ribosomes cell\textsuperscript{-1} \\
Transcript length & $m$ & 300 & codons \\
Ribosome-RBS binding rate & $\alpha_+$ & 0.0001 & ribosome\textsuperscript{-1} RBS\textsuperscript{-1} s\textsuperscript{-1}\\
Ribosome-RBS unbinding rate & $\alpha_-$ & 200 & ribosome-RBS-complex\textsuperscript{-1} s\textsuperscript{-1}\\
\hline
\end{tabular}
\caption{Model parameters used for testing model validity}
\label{tab:model_validation_parameters}
\end{center}
\end{table}

Running a simulation of a single circuit with ribosomes gives a circuit that produces proteins at an average rate of 35.04 proteins per second. This uses an average of 537.97 ribosomes at any point in time which is 2.5\% of all cellular ribosomes. This appears to be the correct order of magnitude since there are approximately 4000 genes, of which perhaps half are active. This gives ~2000 active promoters with approximately 2-4 transcripts per promoter with a total of 4000-8000 cellular transcripts per cell. The 100 transcripts from the synthetic construct constitute 1.25-2.5\% of the total cellular transcripts and therefore we would expect the same proportion of the total cellular ribosomes to be on circuit transcripts.

\section{Modelling control points}
\label{sec:control_points}

We use a simple two circuits simulation to investigate the effect of changing the parameters associated with the different control points. Since we do not know the exact biological parameters for the systems we are investigating we cannot expect an accurate quantitative prediction of the impacts of specific changes. However, we can do a comparative investigation where we look at the qualitative and relative changes in system behaviour when we change the control point parameters.

\subsection{Promoter strength and copy number}

The model being considered in this paper only captures ribosomal availability and therefore when considering the number of circuit transcripts, it is independent of the mechanisms that cause changes in the amount of mRNA. Plasmid backbones are not considered as part of this model due the the higher levels of complexity of different origins of replication and resistance markers. However, a suitable approach for future work would be to characterise the behaviour of the backbone and use this modelling approach to predict how to best optimise the design of the circuit contained in the plasmid given a set of constraints.

A suitable approach would be to characterise the behaviour of the backbone and use this modelling approach to predict how to best optimise the design of the circuit contained in the plasmid given a set of constraints.

\begin{figure}[H]
\includegraphics[width=\textwidth]{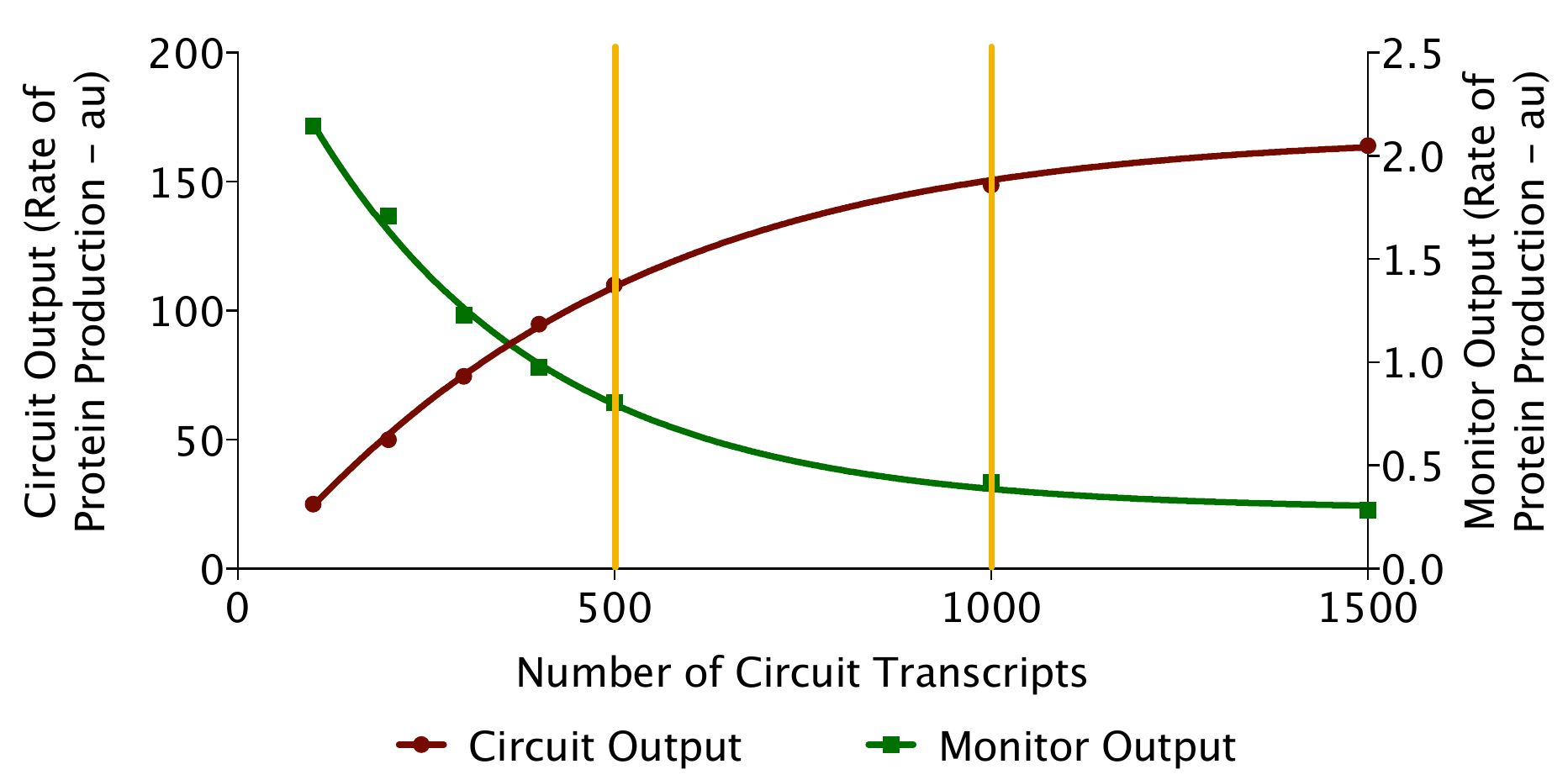}
\caption[Modelled Impact of Transcript Number on Circuit and Monitor Outputs]{Modelled impact of transcript number on circuit and monitor outputs. This figure shows both both monitor output and circuit output for a range of circuit transcript numbers. Lines represent best fit of hill curves using GraphPad Prism with no parameter constraints.}
\label{fig:modelling_transcripts_line}
\end{figure}

In this modelling approach the plasmid copy number and promoter strength are compounded into a single variable - the number of transcripts. Figure \ref{fig:modelling_transcripts_line} shows the amount of circuit output and monitor output for the model system for a range of transcript numbers. At low levels of transcripts ($<400$ per cell) the relationship between transcript number and circuit output is approximately proportional. Similarly, the relationship between the number of transcripts and monitor output is approximately linear in this region. This indicates that for a given number of ribosomes and for transcript numbers in this range, all transcripts use a similar number of ribosomes to produce proteins at a similar rate.

As the number of transcripts increases, the system becomes saturated with respect to transcripts and large increases in the number of transcripts cause relatively small increases and decreases in circuit output and monitor output respectively.

\begin{figure}[H]
\includegraphics[width=\textwidth]{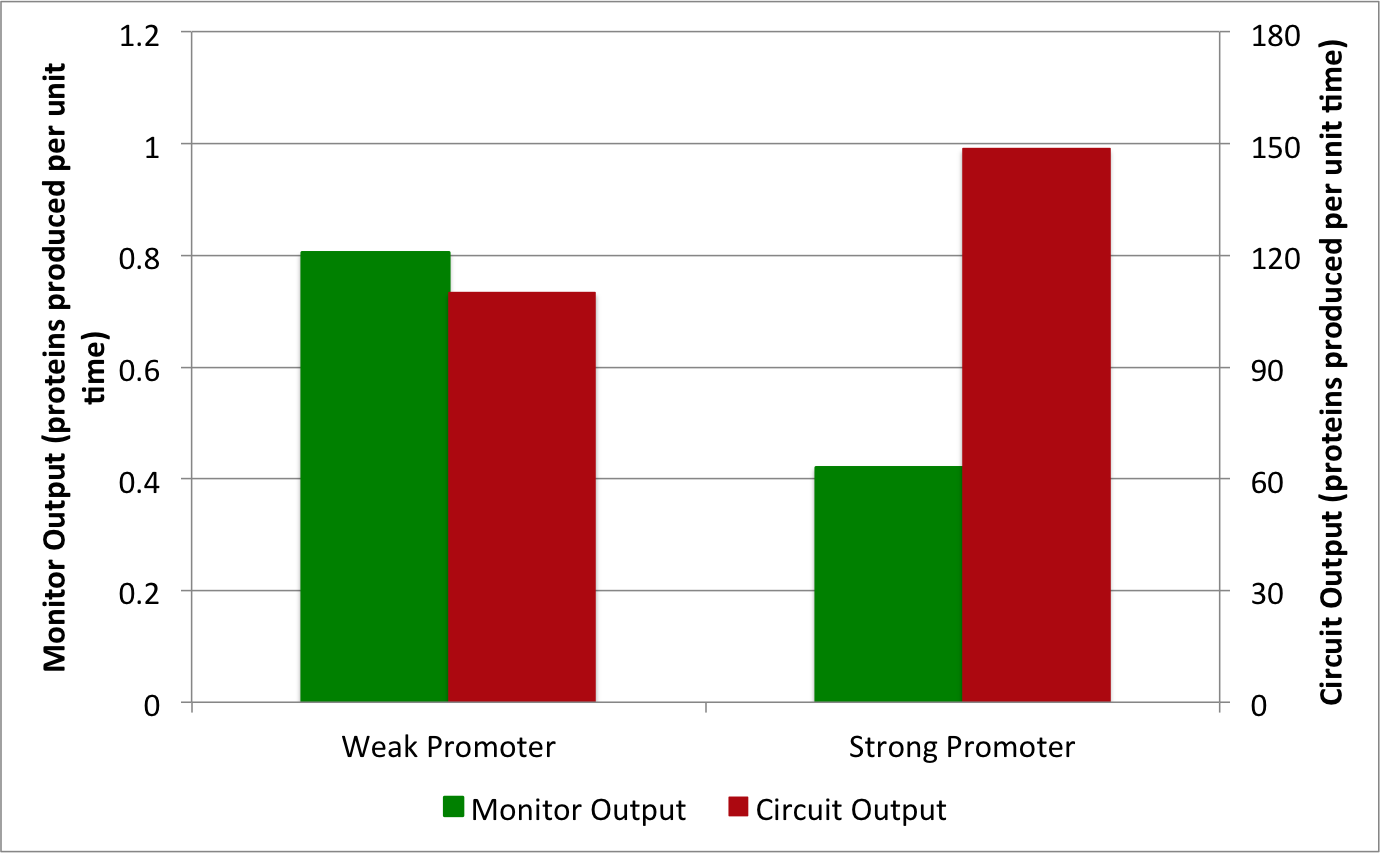}
\caption[Modelled Transcript Number Impact on Circuit and Monitor Outputs]{Modelled impact of transcript number on circuit and monitor output. This figure shows that a higher number of transcripts causes higher circuit output and lower monitor output.}
\label{fig:modelling_transcripts_col}
\end{figure}

The vertical yellow lines in Figure \ref{fig:modelling_transcripts_line} show the time points corresponding to the simulated data given in Figure \ref{fig:modelling_transcripts_col}. 

%

\subsection{RBS strength and codon usage}

We modelled the system with a range of different RBS strengths as well as two different codons usages. The fast codon version has uniform elongation rates of 1 along a 100 codon transcript and the slow codon version has uniform elongation rates of 1 along a 100 codon transcript with the exception of elongation rates of 0.5 for codons 85 to 95.

\begin{figure}[H]
\includegraphics[width=\textwidth]{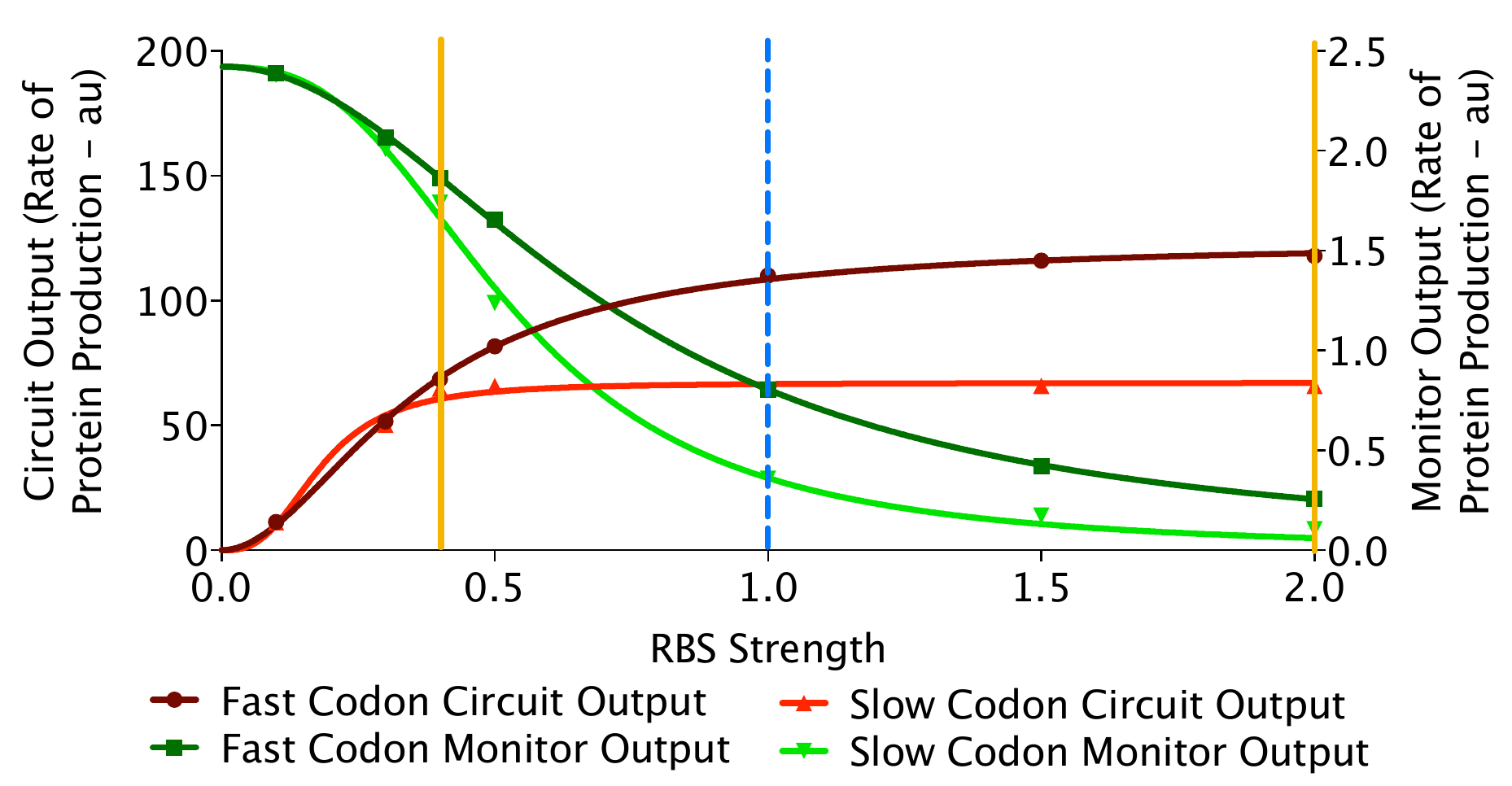}
\caption[Modelled Impact of RBS and Codon Usage on Circuit and Monitor Outputs]{Modelled impact of RBS strength and codon usage on circuit and monitor outputs. Shows both monitor output and circuit output for a range of RBS strengths for two different codon usages. Lines represent best fit of hill curves using GraphPad Prism with no parameter constraints.}
\label{fig:modelling_RBS}
\end{figure}

Both codon usage and RBS have a large impact on the behaviour of the circuit. Figure \ref{fig:modelling_RBS} shows how both codon usage and RBS strength affect the monitor output and the circuit output. For both codon usages, as the RBS strength increases at low levels ($<0.4$) the relationship between circuit output and RBS strength is approximately linear. As the RBS strength continues to increase, the circuit output reaches a saturation level. Slower codons heavily impact the maximum output of the circuit. This is due to slower codons imposing a lower maximum flux of ribosomes through the system. Also, for slower codons this saturation is reached at lower RBS strength. This intuitively makes sense since slower codons will cause a lower maximum flux through the system and a higher rate of recruitment of ribosomes onto the transcript will cause this maximum to be reached.

In terms of monitor output, for RBS strengths $<1$, the relationship between RBS strength and monitor output is approximately linear. For higher RBS strengths, the monitor output tends to a lower asymptote. The slower codon circuit causes a decrease in monitor output.

\begin{figure}[H]
\includegraphics[width=\textwidth]{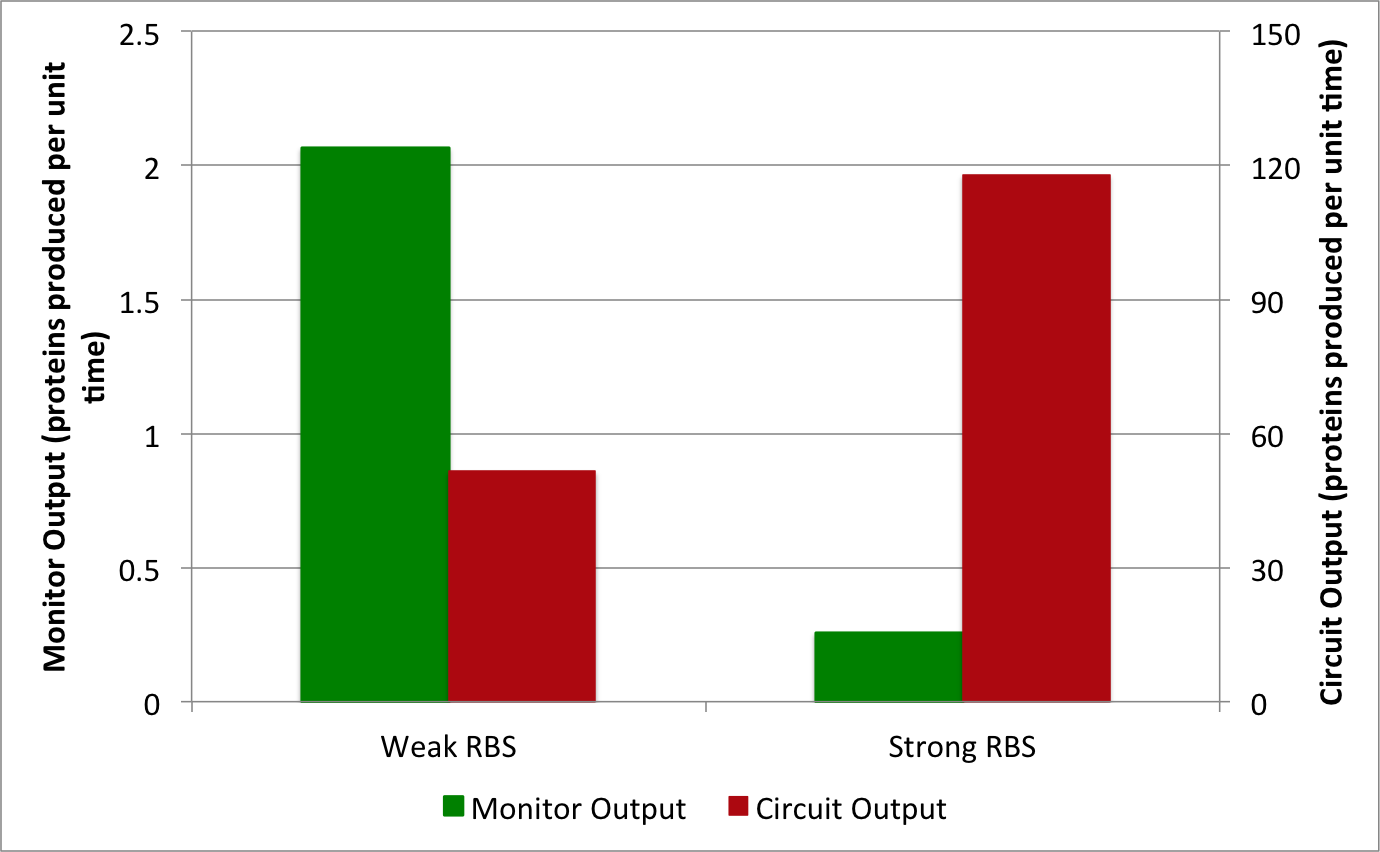}
\caption[Modelled Impact of RBS Strength on Circuit and Monitor Outputs]{Modelled impact of RBS strength on circuit and monitor output shows that a stronger RBS causes higher circuit output and lower monitor output.}
\label{fig:modelling_RBS_columns}
\end{figure}

\begin{figure}[H]
\includegraphics[width=\textwidth]{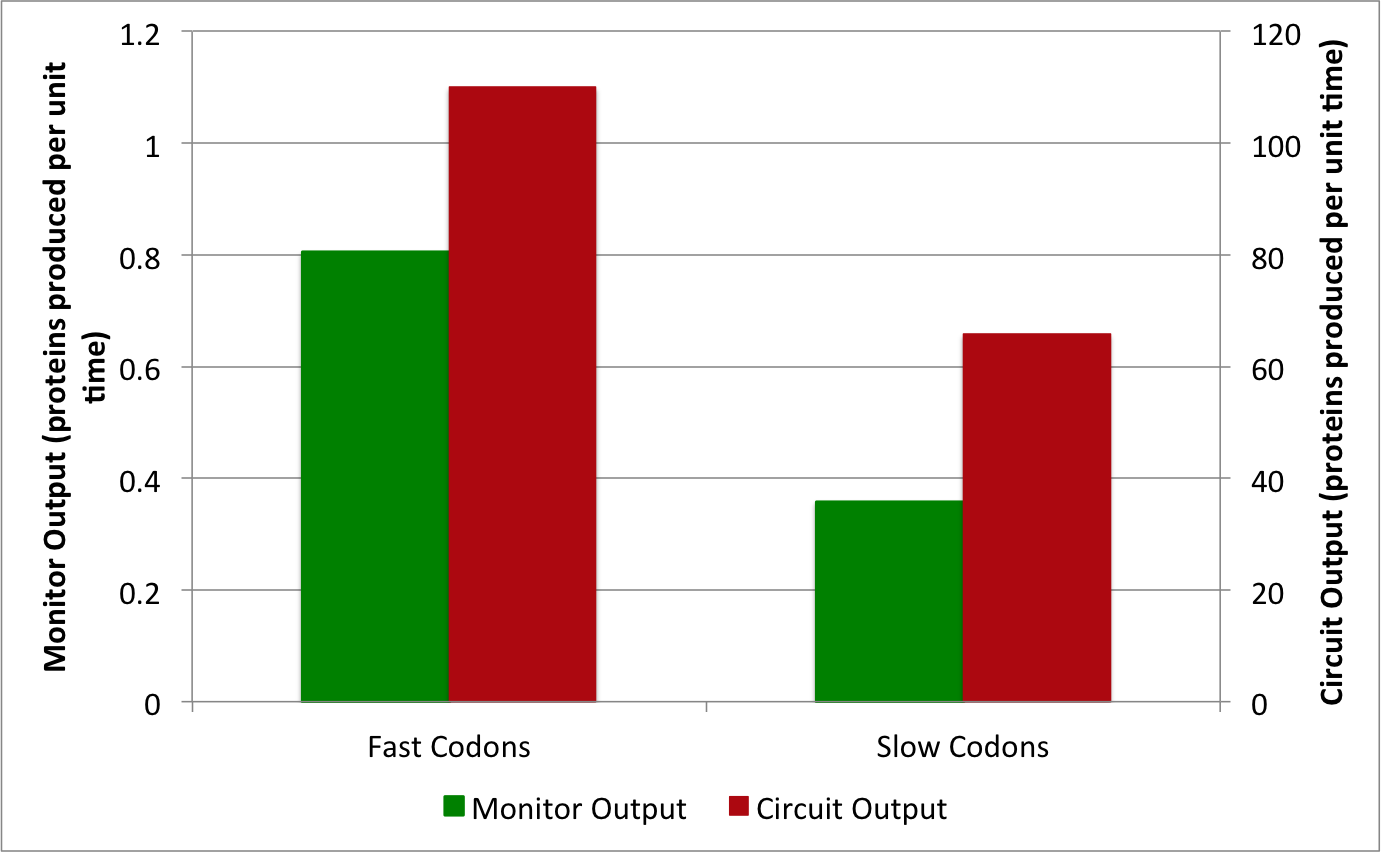}
\caption[Modelled Impact of Codon Usage on Circuit and Monitor Outputs]{Modelled impact of codon usage on circuit and monitor output shows that slower codons in the circuit cause lower circuit output as well as lower monitor output.}
\label{fig:modelling_codon_columns}
\end{figure}

The yellow lines represent the time point at which the data represented in Figure \ref{fig:modelling_RBS_columns} are considered, while the dashed blue line represents the time point at which the data represented in Figure \ref{fig:modelling_codon_columns} are considered.
\\

\begin{rem}
Note that our model is unable to capture the known phenomenon of reduced circuit output at the highest RBS strengths. This is because we are not including cellular response and adaptation in this model, though it would be a very interesting thing to include into the model, which we plan to do in future work.
\end{rem}

%

\section{Obtaining similar circuit output with different burden levels}

In this section we investigate ways to design two circuits with similar circuit output but different monitor output (resource usage). 

This can done by changing both the RBS and promoter together so that in one case a stronger promoter is used with a weaker RBS and in the other case a weaker promoter was used with a stronger RBS.

We simulated this situation using RBS strengths and transcript numbers that were above the regions where we saw a proportional behaviour between the variables and circuit output. The weak RBS had strength 0.4 and the strong RBS had strength of 2, a 5-fold difference. The number of transcripts used was 300 for the weak promoter and 500 for the strong promoter.

\begin{figure}[H]
\includegraphics[width=\textwidth]{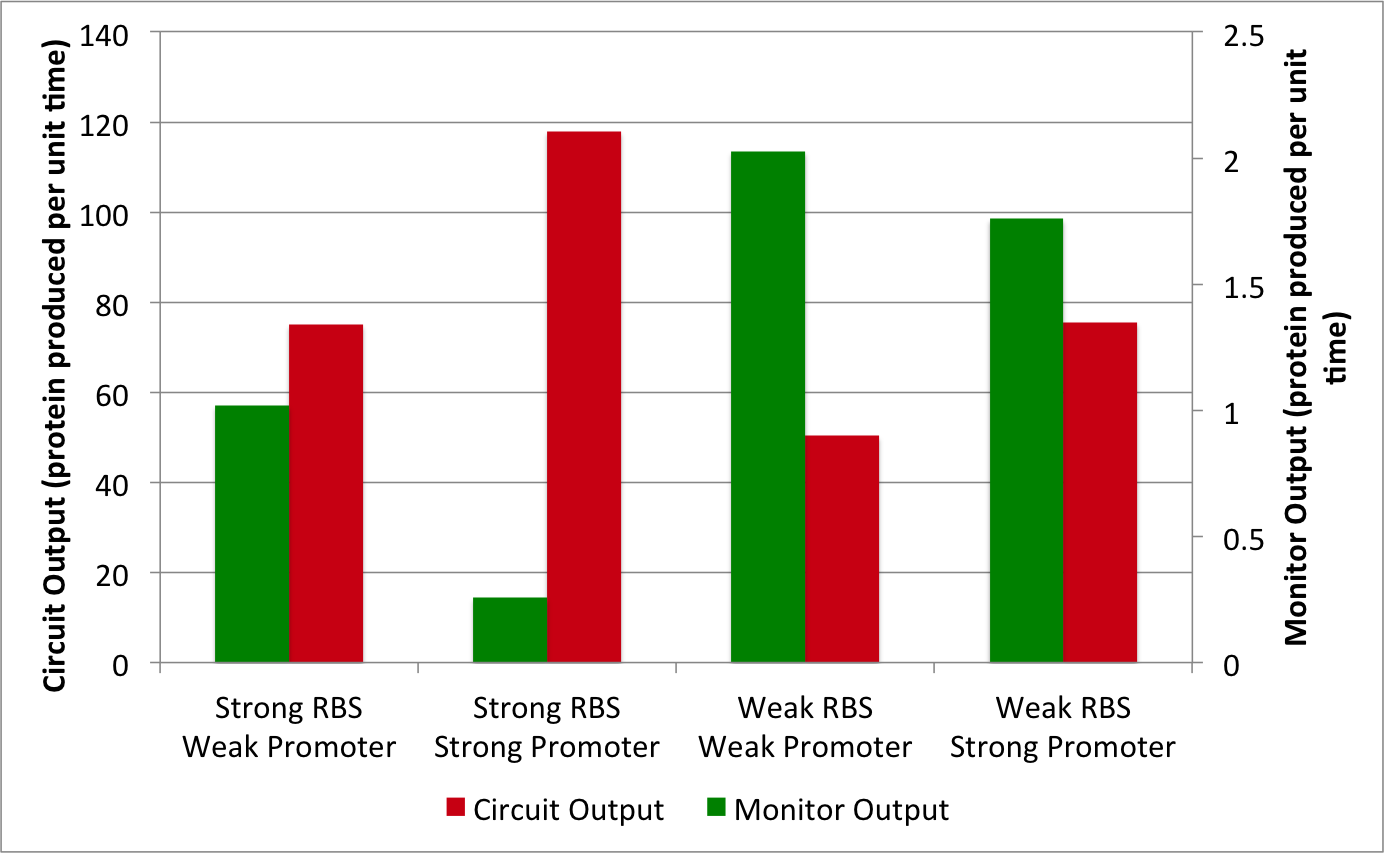}
\caption[Obtaining Similar Circuit Output with Different Burden Levels - Modelling]{Obtaining Similar Circuit Output with Different Burden Levels - Modelling}
\label{fig:modelling_sameoutput_differentburden_columns}
\end{figure}

Figure \ref{fig:modelling_sameoutput_differentburden_columns} shows the data obtained from the modelling and shows that the circuit output for the strong promoter/weak RBS and weak promoter/strong RBS is approximately the same, however the output from the monitor is higher for the strong promoter/weak RBS version. The weak promoter/weak RBS construct has the lowest circuit output and highest monitor output whilst the strong promoter/weak RBS construct has the highest circuit output and lowest monitor output.


\section{Conclusion}

In this paper we have shown the development of a model of translation. This model has been designed so that it is able to incorporate the effect of codon usage on gene expression as well as ribosomal usage. This model assumes that the competition for transcriptional resources is less important than the competition for translational resources and can be neglected.

We used a random walk approach to model the behaviour of ribosomes within a cell moving from a free pool and reversibly binding to a transcript before moving unidirectionally along the transcript. This approach was then developed into a deterministic steady-state model using expectations. We then proved that there was a unique solution to these equations. However, unfortunately for any realistic circuit we might want to model these equations are not analytically solvable and therefore we must use numerical methods to simulate their behaviour.

We provide a python script that is able to simulate a cell with an arbitrary number of mRNA species where the length, codon speed, RBS strength and number of transcripts call all be defined. We then ran a simulation of this with biologically realistic numbers and obtained outputs that were within realistic bounds.

This model was subsequently used to predict the impact of changing the number of transcripts (to reflect a change in copy number or promoter strength). These results showed that there are diminishing returns for protein production levels as transcript numbers are increased and that both monitor output and circuit output tend towards asymptotes.

The model also predicted that increases in RBS strength would lead to saturating increases in circuit output as well as decreases in monitor output. Since the model did not include any cellular feedback we were not able to observe decreased circuit outputs for particularly high RBS strengths. We also observed that introducing slow codons into the transcript caused a decrease in both monitor output and circuit output, reinforcing the value of codon optimization.

Crucially, the model was also able to reflect the ability for two circuits to have the same circuit output whilst causing different levels of burden. A construct with low transcript numbers and high RBS strength (weak promoter, strong RBS) was shown to cause a higher level of burden than a circuit with a higher number of transcripts (stronger promoter) and weaker RBS that gave the same circuit output. This shows that our model may be used to uncover non-intuitive host-circuit behaviours.


\addcontentsline{toc}{chapter}{Bibliography}
\bibliographystyle{hunsrt}
\bibliography{references.bib}

\begin{thebibliography}{10}

\bibitem{scottinterdependence2010}
Matthew Scott, Carl Gunderson, Eduard Mateescu, Zhongge Zhang, and Terence Hwa.
\newblock Interdependence of cell growth and gene expression: origins and
  consequences.
\newblock {\em Science {(New} York, {N.Y.)}}, 330(6007):1099--1102, 2010.

\bibitem{perettimechanistically1986}
S~Peretti and J~Bailey.
\newblock Mechanistically detailed model of cellular metabolism for
  glucose-limited growth of escherichia coli {B/r-A.}
\newblock {\em Biotechnology and bioengineering}, 28(11):1672--1689, 1986.

\bibitem{perettisimulations1987}
S~Peretti and J~Bailey.
\newblock Simulations of host-plasmid interactions in escherichia coli: Copy
  number, promoter strength, and ribosome binding site strength effects on
  metabolic activity and plasmid gene expression.
\newblock {\em Biotechnology and bioengineering}, 29(3):316--328, 1987.

\bibitem{levycoordination2009}
Sagi Levy and Naama Barkai.
\newblock Coordination of gene expression with growth rate: a feedback or a
  feed-forward strategy?
\newblock {\em {FEBS} letters}, 583(24):3974--3978, 2009.

\bibitem{marchisiocomputational2008}
M~Marchisio and J~Stelling.
\newblock Computational design of synthetic gene circuits with composable
  parts.
\newblock {\em Bioinformatics {(Oxford,} England)}, 24(17):1903--1910, 2008.

\bibitem{klumppgrowth2009}
Stefan Klumpp, Zhongge Zhang, and Terence Hwa.
\newblock Growth rate-dependent global effects on gene expression in bacteria.
\newblock {\em Cell}, 139(7):1366--1375, 2009.

\bibitem{tanemergent2009}
Cheemeng Tan, Philippe Marguet, and Lingchong You.
\newblock Emergent bistability by a growth-modulating positive feedback
  circuit.
\newblock {\em Nature chemical biology}, 5(11):842--848, 2009.

\bibitem{mitarairibosome2008}
Namiko Mitarai, Kim Sneppen, and Steen Pedersen.
\newblock Ribosome collisions and translation efficiency: optimization by codon
  usage and {mRNA} destabilization.
\newblock {\em Journal of molecular biology}, 382(1):236--245, 2008.

\bibitem{vindsynthesis1993}
J~Vind, M~Sørensen, M~Rasmussen, and S~Pedersen.
\newblock Synthesis of proteins in escherichia coli is limited by the
  concentration of free ribosomes. expression from reporter genes does not
  always reflect functional {mRNA} levels.
\newblock {\em Journal of molecular biology}, 231(3):678--688, 1993.

\bibitem{de_voshow2011}
Dirk De~Vos, Frank Bruggeman, Hans Westerhoff, and Barbara Bakker.
\newblock How molecular competition influences fluxes in gene expression
  networks.
\newblock {\em {PloS} one}, 6(12), 2011.

\bibitem{chouclustered2004}
Tom Chou and Greg Lakatos.
\newblock Clustered bottlenecks in {mRNA} translation and protein synthesis.
\newblock {\em Physical review letters}, 93(19), 2004.

\bibitem{basutraffic2007}
Aakash Basu and Debashish Chowdhury.
\newblock Traffic of interacting ribosomes: Effects of single-machine
  mechanochemistry on protein synthesis.
\newblock {\em Physical Review E}, 75, 2007.

\bibitem{macdonaldconcerning1969}
Carolyn~T. {MacDonald} and Julian~H. Gibbs.
\newblock Concerning the kinetics of polypeptide synthesis on polyribosomes.
\newblock {\em Biopolymers}, 7, 1969.

\bibitem{proshkincooperation2010}
Sergey Proshkin, A~Rahmouni, Alexander Mironov, and Evgeny Nudler.
\newblock Cooperation between translating ribosomes and {RNA} polymerase in
  transcription elongation.
\newblock {\em Science {(New} York, {N.Y.)}}, 328(5977):504--508, 2010.

\bibitem{bionumbers}
Bionumbers.

\end{thebibliography}

\end{document}